\documentclass[12pt]{article}
\usepackage{authblk}
\usepackage{geometry}
\usepackage{amsthm}
\usepackage{mathrsfs}
\usepackage{subfig}
\usepackage{caption}
\usepackage{customcommands}
\usepackage{bm}
\usepackage{Nettastyle}
\usepackage{amsmath}
\usepackage{amssymb}
\usepackage{amsfonts}
\usepackage{amsthm}
\usepackage{braket}
\usepackage[normalem]{ulem}
\usepackage{color}
\usepackage{ulem}
\usepackage{mathtools}
\usepackage{tensor} 

\DeclareMathOperator{\tr}{tr}

\theoremstyle{plain}
\newtheorem*{thm-restate}{Theorem \ref{thm:qms_exact}}

\newcommand{\bfig} {\begin{figure}\begin{center}}
\newcommand{\efig}{\end{center}\end{figure}}
\newcommand{\bi}{\begin{itemize}}
\newcommand{\ei}{\end{itemize}}

\newcommand{\Hs}{\mathcal{H}_s}
\newcommand{\Hr}{\mathcal{H}_r}
\newcommand{\Hss}{\mathcal{H}_{s^{\prime}}}
\newcommand{\Hrr}{\mathcal{H}_{r^{\prime}}}
\newcommand{\HR}{\mathcal{H}_R}

\newcommand{\HB}{\mathcal{H}_B}

\newcommand{\Hf}{\mathcal{H}_f}
\newcommand{\HP}{\mathcal{H}_P}

\newcommand{\ran}{\rangle}
\newcommand{\lan}{\langle}

\title{Reflected entropy in an evaporating black hole through non-isometric map}

\author[1,2,3]{Bin Chen,}
\author[2]{Zhi-jun Yin}
\affiliation[1]{Institute of Fundamental Physics and Quantum Technology,\\ \&  School of Physical Science and Technology, \\ Ningbo University, Ningbo, Zhejiang 315211, China}
\affiliation[2]{School of Physics, Peking University, No.5 Yiheyuan Rd, Beijing
		100871, P.R. China}
\affiliation[3]{Center for High Energy Physics, Peking University,
		No.5 Yiheyuan Rd, Beijing 100871, P. R. China}
\emailAdd{chenbin1@nbu.edu.cn}
\emailAdd{yinzhijun@stu.pku.edu.cn}
\abstract{
The black hole information paradox has been an important problem in quantum gravity. In the study of evaporating black hole, it has been proposed that the holographic map between the semi-classical effective description in bulk and the fundamental description in boundary cannot be isometric. In this work, we would like to study the reflected entropy in an evaporating black hole model through non-isometric holographic map. We assume that the evaporating is slowly enough that  it makes sense to ascribe a slowly varying temperature to the Hawking radiation. We then introduce a two-sided black hole model to canonically purify the semi-classical state. The holographic map to the fundamental description is non-isometric and defined by a Haar random unitary matrix.  We show that the  entropy of  radiation in the model  computed in the fundamental description matches the result read from the quantum extremal surface formula and agrees with the Page curve. Furthermore, through non-isometric holographic map, we study the reflected entropies between different regions, including the one between the black holes on different sides, the one between the radiations distributed symmetrically but disconnectedly, and the one between the black hole and the radiation on single side. Our results are consistent with the existing ones based on the effective descriptions.
}

\begin{document}
\maketitle

\section{Introduction}\label{sec:intro}
One important implication of AdS/CFT correspondence  \cite{Maldacena:1997re} is that  the black hole evaporation process must keep the unitarity. The  entropy of Hawking radiation is an important physical quantity for detecting whether black hole evaporation is unitary or not. If black hole evaporation is unitary, the entropy of Hawking radiation should  follow the Page curve \cite{Page:1993wv,Page:2013dx}. In the AdS/CFT correspondence, the boundary entropy and its bulk dual, the quantum extremal surface (QES) formula \cite{Engelhardt:2014gca}, when applied to evaporating black holes, will lead to the so-called ``island rule" \cite{Penington:2019npb,Almheiri:2019psf,Almheiri:2019hni,Almheiri:2019yqk,Penington:2019kki,Almheiri:2019qdq,Geng:2024xpj},  which helps to show that the entropy of Hawking radiation  obeys the Page curve.

The quantum extremal surface has also been used to study the problem of bulk reconstruction, which involves representing bulk operators in the Hilbert space of the boundary CFT \cite{Banks:1998dd,Hamilton:2006az,Heemskerk:2012mn,Bousso:2012mh,Czech:2012bh,Wall:2012uf,Headrick:2014cta}. During the research, it was found that the problem of bulk reconstruction can be discussed within the framework of quantum error correction \cite{Almheiri:2014lwa,Pastawski:2015qua,Dong:2016eik,Harlow:2016vwg}.  The underling key points are as follows. The Hilbert space of the low-energy effective field theory in the bulk can be mapped to the boundary Hilbert space by an approximate isometric map $V: \mathcal{H}_{bulk} \to \mathcal{H}_{boundary}$ (an isometry is a linear map satisfying $V^\dag V = I$). In the language of quantum error correction, $\mathcal{H}_{bulk}$ is the ``logical" Hilbert space, and $\mathcal{H}_{boundary}$ is the ``physical" Hilbert space. If the boundary Hilbert space can be tensor factorized into two parts $\mathcal{H}_{boundary} = \mathcal{H}_A \otimes \mathcal{H}_{\bar{A}}$, then the bulk can be decomposed along the corresponding quantum extremal surface$X_A^{min}$ into $\mathcal{H}_{bulk}  = \mathcal{H}_a \otimes \mathcal{H}_{\bar{a}}$. In this case, for the bulk density operator $\rho$, we have the QES formula:
\begin{equation}
S(\tr_{\bar{A}}(V \rho V^\dagger)) \approx \frac{\mathrm{Area}(X^{\min}_A)}{4} + S(\rho_a).
\end{equation}
Strictly speaking, the above bulk reconstruction only applies to the emergence of spacetime outside the blackhole horizon. 

\begin{figure}
    \centering
    \includegraphics[width=0.75\linewidth]{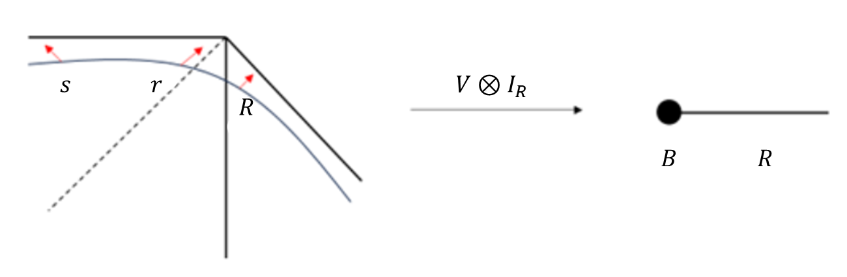}
    \caption{Holographic map  of a single-sided black hole coupled with the radiation $R$. The  $s$ modes, the $r$ modes, and the $R$ modes in the radiation region appear on the selected Cauchy surface in the effective description.}
    \label{fig:enter-label}
\end{figure}

The above holographic map encounters serious problems in describing the spacetime behind the blackhole horizon. For example, in the late stage of black hole evaporation,  the degrees of freedom in the bulk effective field theory will vastly exceed the degrees of freedom on the boundary. On the one hand, as a simple model, one may take the black hole as a quantum system $B$ with Hilbert space $\mathcal{H}_B$ containing the microscopic degrees of freedom of black hole, coupled to a ``reservoir" $R$ with Hilbert space $\mathcal{H}_R$. The Hawking radiation from the black hole can propagate out into the reservoir. In this so-called {\it fundamental} description, the quantum system could belong to a high-energy band of holographic CFT, and the reservoir $R$ could be a kind of free-field theory on a half-space in one higher dimension than the holographic CFT if we may ignore the interaction of radiations safely. On the other hand,  from the bulk effective field theory perspective, there are two kinds of modes inside the black hole: one kind of modes consists of the states falling into the black hole, denoted as $s$, and the other kind of ones denoted as $r$ are entangled with the Hawking radiations. The semi-classical bulk description is referred to as the {\it effective} description. The holographic map in this case is defined as
\begin{equation}
V\otimes I_R: \mathcal{H}_s \otimes \mathcal{H}_r\otimes \mathcal{H}_R \rightarrow \mathcal{H}_B\otimes \mathcal{H}_R,
\end{equation}
as shown in Fig. \ref{fig:enter-label}. Here the interactions between two kinds of modes have been neglected.
At the late stage of black hole evaporation, the entanglement entropy of the radiations in the effective  description far exceeds the entropy of the black hole, 
\begin{equation}
S(\rho_r) = S(\rho_R) \gg \log |B|. 
\end{equation}
In other words, the dimension of the Hilbert space $\mathcal{H}_r$ is much larger than that of $\mathcal{H}_B$, $|r|\gg |B|$. Therefore, the linear map $V:\mathcal{H}_s \otimes \mathcal{H}_r\mapsto \mathcal{H}_B$ must be non-isometric, and a large number of null states in the effective description will be annihilated by the map. In \cite{Akers:2022qdl}, the authors constructed a non-isometric holographic map from the effective description to the fundamental description and proved that it has properties such that observing the null states requires the operations of exponential complexity. It was showed that these non-isometric codes protected by computational complexity could explain the emergence of the black hole interior.


The model in  \cite{Akers:2022qdl} was refined  in \cite{Gyongyosi:2023sue}  by taking into account of the energy conservation and the thermal nature of the Hawking radiation.  In \cite{Gyongyosi:2023sue}, it was assumed that the black hole evaporates slowly enough that it makes sense to ascribe a slowly varying temperature to the Hawking radiation. In this so-called quasi-adiabatic regime, Hawking's derivation is valid, and it requires that $S_{BH}\gg c$, where $c$ is the degrees of freedom of massless modes. More precisely, the black hole is assumed to evaporate in a series of time steps whose sizes are of
the order of the scrambling time of the black hole. In this picture, the black hole evolves approximately through a sequence of microcanonical states, and the semi-classical state is now a thermofield double with a slowly varying temperature. One may canonically purify the thermofield double state by considering the two-sided black hole. In this setup, one may discuss richer physics\cite{Maldacena:2001kr,Almheiri:2019yqk,Mathur:2014dia}. Now the entire system is naturally divided into multiple regions, we can study the entanglements between these regions. In this work, we plan to study the reflected entropy in this two-sided black hole model.     

The reflected entropy was first introduced in the study of canonical purification for the entanglement wedge cross-section\cite{Dutta:2019gen}, and has been under intensive study in various situations, including holographic models\cite{Bao:2019zqc,Jeong:2019xdr,Chandrasekaran:2020qtn,Li:2020ceg,Basu:2021awn,Li:2021dmf,Ling:2021vxe,Akers:2022max,Basu:2022nds,Basak:2022cjs,Chen:2022fte,Vasli:2022kfu,Lu:2022cgq,Afrasiar:2022fid,Afrasiar:2023jrj,Basu:2023jtf,Basak:2023bnc,Basu:2024bal,Ahn:2024gep,Yuan:2024yfg}, random tensor networks\cite{Akers:2021pvd,Akers:2022zxr,Czech:2023rbh,Akers:2024pgq}, the free scalar and fermion field theories\cite{Bueno:2020vnx,Bueno:2020fle,Dutta:2022kge,Basak:2023uix}, Chern-Simons theory\cite{Berthiere:2020ihq,Liu:2021ctk,Sohal:2023hst}, conformal field theory\cite{Kusuki:2019rbk,Kudler-Flam:2020url,Moosa:2020vcs,Kudler-Flam:2020xqu,Berthiere:2023gkx}, Lifshitz theory\cite{Berthiere:2023bwn}, etc. The reflected entropy is a probe of tripartite entanglement \cite{Akers:2019gcv,Hayden:2021gno}. It is remarkable that even though the monotonicity of the reflected entropy gets lost for certain quantum state\cite{Hayden:2023yij} and for the ground state in free Lifshitz field theories\cite{Berthiere:2023bwn}, it holds for holographic state\cite{Dutta:2019gen}, as the consequence of the “entanglement
wedge nesting” property.  

We are going to compute the reflected entropies between different regions in the two-sided black hole model. Our study  is in spirit similar to the one appearing in the 2D eternal black hole plus CFT model \cite{Li:2020ceg,Chandrasekaran:2020qtn}, where the two-sided black hole was used to discuss the thermal equilibrium between the black hole and the thermal bath  \cite{Almheiri:2019yqk}.  We will calculate the reflected entropy between the black holes on different sides,  the one between the radiations distributed in a symmetric but disconnected way,  and the one between black hole and the radiations. In the case at hand, we verify the equality between the reflected entropy between the black hole and the radiation in the single-sided black hole model and the entanglement entropy of radiations in the two-sided black hole model, despite the  fact that  two  models  have different holographic maps. To obtain analytical results, we focus on the high-temperature limit, where the system can be taken as the microcanonical ensemble. In this case, we can clearly observe the changes in reflected entropy during the black hole evaporation process.

The remaining parts of the paper are organized as follows. In Section 2, we introduce briefly the two-sided black hole. In Section3, we compute the reflected entropies and the mutual information between different regions. We end with conclusion and discussion in Section 4.


\section{A model for two-sided black hole}\label{sec:model}

We first review briefly the model in \cite{Akers:2022qdl}. In this model, to achieve the non-isometry of the holographic map, we consider a larger Hilbert space $\Hs \otimes \Hr\otimes \Hf$, where $f$ are some extra degrees of freedom such that:
\begin{equation}
	|f||s||r|=|P||B|.
\end{equation}
For a positive integer $|P|$, there is a tensor decomposition of the Hilbert space
\begin{equation}\label{Hilbert}
	\Hs \otimes \Hr\otimes \Hf=\HB\otimes\HP.
\end{equation}
The holographic map is defined as
\begin{equation}
	V\equiv \sqrt{|P|}\, \lan 0|_P U |\psi_0\ran_f,
\end{equation}
where $|\psi_0\ran_f$ is a fixed quantum state on $\Hf$, 
$|0\ran_P$ is a fixed quantum state on $\HP$, and $U$ is a unitary operator on $\Hs \otimes \Hr\otimes \Hf$, which satisfies the statistical properties of being Haar random. The coefficient $\sqrt{|P|}$ is used for normalization. We can regard $|\psi_0\ran_f$ as describing any extra effective-field-theory degrees of freedom we are not interested in. This model achieves the non-isometry of the holographic map through the post-selection on the unitary operator by $\lan 0|_P $.
\begin{figure} 
	\centering
	\includegraphics[width=0.5\linewidth]{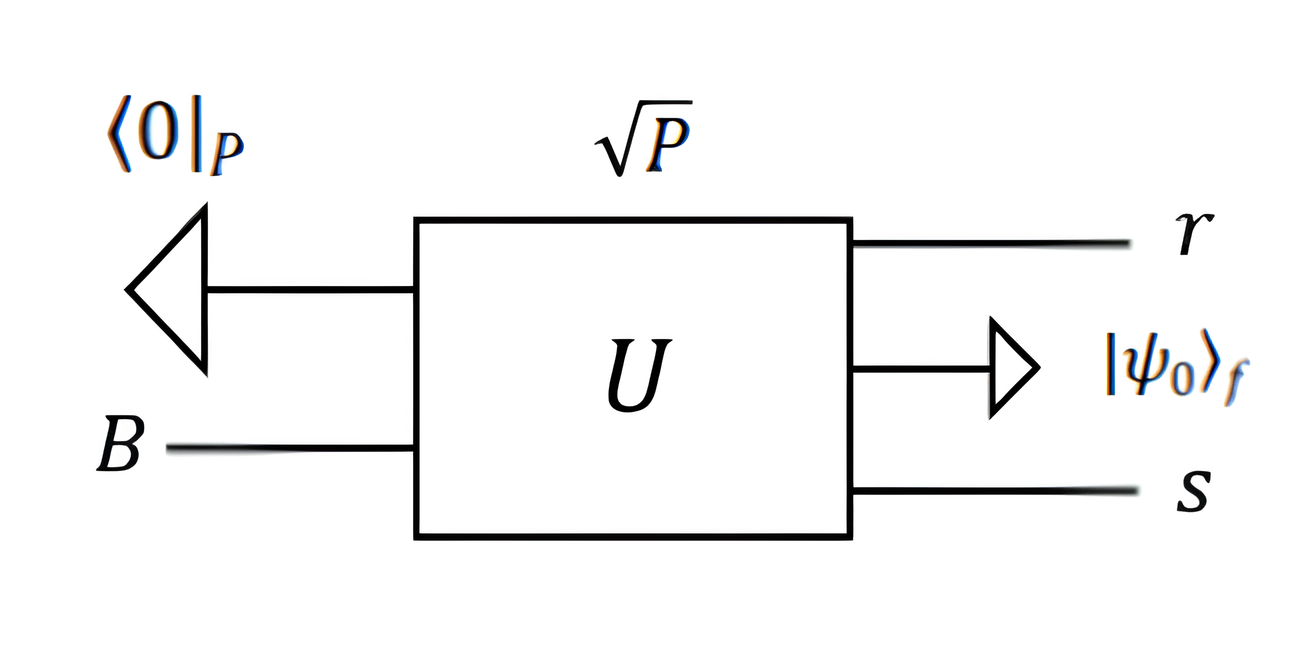}
	\caption{The diagram of the holographic map $V$,  in which  $U$ is a unitary operator and the external legs on both sides correspond to the degrees of freedom of the respective Hilbert spaces. The degrees of freedom of $f$ and $P$ are fixed by $|\psi_0\rangle_f$ and $\langle 0|_P$. For clarity, the order of the external leg indices in the diagram matches the indices of $U$ in (\ref{int}), i.e., the left side corresponds to the index $i$, and the right side corresponds to the index $j$.}
	\label{fig:v}
\end{figure}

For an evaporating black hole, the $r$ modes form  maximally entangled pairs with the radiations \cite{Mathur:2009hf}. Ignoring the contribution of $|\psi_0\ran_f$, we consider that the black hole is formed by the infalling 
$s$ state. From this,we can find that $|P|=|r||R|$,  $|0\ran_P=|\mathrm{MAX}\ran_{rR}$ is the maximal entangled state on $\Hr\otimes\HR$\cite{Akers:2022qdl,Gyongyosi:2023sue}. So we can derive the form of the holographic map: 
\begin{equation}\label{V}
	V= \sqrt{|r||R|}\, \lan \mathrm{MAX}|_{rR} U.
\end{equation}

\subsection{Two-sided black hole model}

Next, let us discuss the case of a two-sided black hole. Typically, in AdS/CFT,  without matter and evaporation, a black hole is dual to a mixed state in the CFT. In the simplest setting, a static black hole in AdS is dual to a thermal state in CFT. And a two-sided black hole is dual to the purification of this mixed state, the so-called thermal-field double state, in two copies of the conformal field theory\cite{Maldacena:2001kr}. Here we generalize the picture to take into account the radiations. We assume that the radiation is slow enough such that we can approximately take the black hole as a state in (quasi)-thermal equilibrium. In our model, the infalling state $s$ under the effective description  corresponds to the formation process of the black hole. We assume it to be varying very slowly as well such that it does not change the mass of the black hole quickly. It could be a mixed state or a pure state.  In any case,  we may have a copy $s^\prime$ of the infalling state $s$ to purify $s$, with the pure state composed of $s$ and $s^\prime$ being denoted as $|\mathrm{S}\rangle_{ss^\prime}$. 

\begin{figure}[t]
	\centering
	\includegraphics[width=1.2\linewidth]{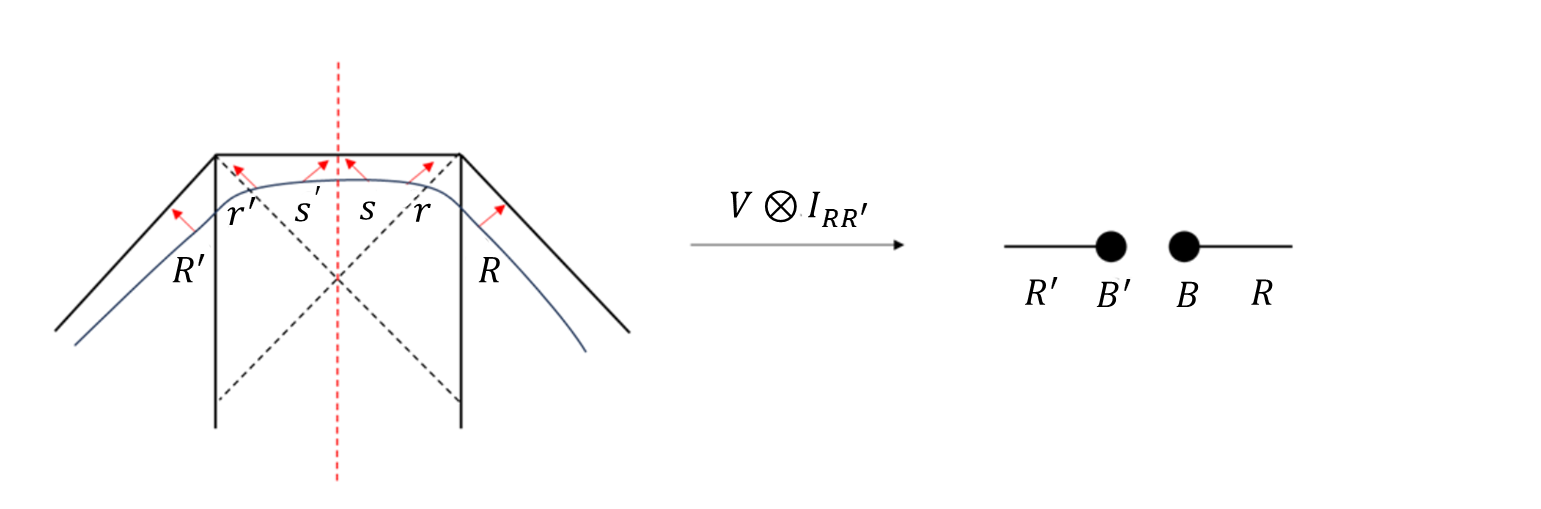}
	\caption{The holographic map of the model in which a two-sided black hole is coupled with the radiations on two sides. On the left side, the effective description is obtained by doubly copying a single-sided evaporating black hole along the red dashed line , resulting in a doubled system. On the right side, the basic description corresponds to the double copies of  the quantum system representing the black hole region coupled with the radiation.  For the holographic map, the mapping on the radiation regions $R$ and its copy $R^\prime$ are trivial. Similar diagram has been used in the discussions of JT gravity with the radiations at a finite temperature,  as shown in \cite{Almheiri:2019yqk}.}
	\label{fig:vtil}
\end{figure}

We consider the coupling of a two-sided black hole with the radiations $R$ and $R^\prime$ on two sides. In the fundamental description, we have to doubly copy the CFT and the reservior, as shown in Fig. \ref{fig:vtil}. At this point, the modes $r$ and $r^\prime$ form maximally entangled states with the modes $R$ and $R^\prime$, respectively, denoted as $|\mathrm{X}\rangle_{rR}$ and $|\mathrm{X^\prime}\rangle_{r^\prime R^\prime}$. In the effective description, the quantum state of the system is taken as
\begin{equation}
	|\psi\rangle = |\mathrm{X}\rangle_{rR} |\mathrm{X^\prime}\rangle_{r^\prime R^\prime} |\mathrm{S}\rangle_{ss^\prime} .
\end{equation}
Now the holographic map is as shown in figure \ref{fig:vtil}. If we regard $s$ and $s^\prime$, $r$ and $r^\prime$, $B$ and $B^\prime$ respectively as the double copy  of the former Hilbert space in (\ref{Hilbert}),  following the analysis before, we can define the holographic map as
\begin{equation}\label{Vtilmap}
	\tilde{V} \equiv |X|\langle \mathrm{MAX} |_{rR} \langle \mathrm{MAX}|_{r^\prime R^\prime} \tilde{U} 
\end{equation}
where $ \tilde{U} $ is a unitary operator on $\Hs \otimes\Hss\otimes\Hr\otimes\Hrr$, which satisfies the statistical properties of being Haar random. For convenience, we set $|X|=|r||r^\prime|=|R||R^\prime|$ and $|Y|=|B||B^\prime|$.

\subsection{Averaging and the Haar integral}

In the following, we will show how to compute the  entropy of radiations in the fundamental description. 
In our model, for a holographic map $\tilde{V}$, the corresponding unitary operator $\tilde{U}$ is not a random variable but a fixed value. However, when calculating some quantities with small fluctuations, we can average over $\tilde{U}$ to simplify the computation. Such quantities are referred to as the self-averaging quantities. The Rényi entropy we are going to calculate is a self-averaging quantity. After averaging, we may compute  the von Neumann entropy  by using the formula
\begin{equation}
	S(\rho) = \lim_{n \to 1} \frac{1}{1 - n} \log \left( \tr\left(\overline{\rho^n}\right)\right). 
\end{equation}

In the calculation, we need to take the average under the Haar measure. The integral under the Haar measure can be generally written as $\int dUf\left(U, U^\dagger\right)$ where $U$ and its conjugate transpose $U^\dagger$ are elements of the unitary group $U(N)$, and $dU$ is the Haar measure, satisfying the normalization condition: $\int dU = 1$.  In order to obtain non-vanishing integrals under the Haar measure, we need to have the same number of $U$ and $U^\dagger$. For such integrals, we have:
\begin{equation}\label{int}
	\int dU U_{i_1 j_1} \ldots U_{i_n j_n} {U^\dag}_{j_1^\prime i_1^\prime} \ldots {U^\dag}_{j_n^\prime i_n^\prime} = \sum_{\sigma, \tau \in S_n} \delta_{i_1 i_{\sigma(1)}^\prime} \ldots \delta_{i_n i_{\sigma(n)}^\prime} \delta_{j_1 j_{\tau(1)}^\prime} \ldots \delta_{j_n j_{\tau(n)}^\prime} \mathrm{Wg}(\sigma \tau^{-1}, N)
\end{equation}
where $\mathrm{Wg}(\sigma \tau^{-1}, N)$ is the Weingarten function\cite{Collins:2003ncs,Collins:2006jgn}, and $\sigma$ and $\tau$ are permutations in the symmetric group $S_n$.

\begin{figure}
	\centering
	\includegraphics[width=0.9\linewidth]{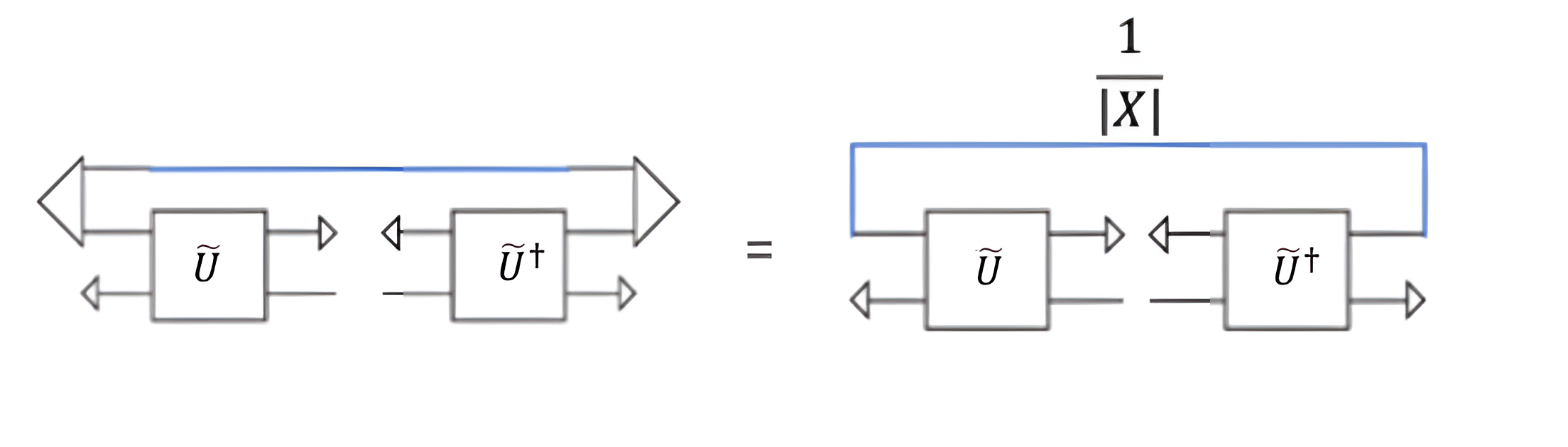}
	\caption{Simplification of the diagram: the blue line represents the  index contraction, and the large triangle to the left of the blue line represents the state $|\mathrm{X^\prime}\rangle_{r^\prime R^\prime}$; after index contraction, the  left diagram is simplified to the right figure with an additional coefficient $\frac{1}{|X|}$.}
	\label{fig:simp}
\end{figure}

When calculating the $n$-th Rényi entropy, if $n$ is large, the Weingarten function becomes complex. However, for our model, we usually assume that the degrees of freedom of the black hole and the radiations are large, thus we can assume $N =|X|^2|Y|\gg 1$. In this case, we can approximate the Weingarten function as
\begin{equation}
	\mathrm{Wg}(\pi, N) = \frac{\delta(\pi, e)}{ N^{n}} + O\left(\frac{1}{N^{n+1}}\right)
\end{equation}
where $\delta(\pi, e)$ is the delta function on the group elements in the symmetric group $S_n$, and $e$ is the identity element. Thus, we can approximate the equation (\ref{int}) as
\begin{equation}\label{approxint}
	\int dU U_{i_1 j_1} \ldots U_{i_n j_n} {U^\dag}_{j_1^\prime i_1^\prime} \ldots {U^\dag}_{j_n^\prime i_n^\prime} = \sum_{\sigma\in S_n} \frac{1}{N^n} \delta_{i_1 i_{\sigma(1)}^\prime} \ldots \delta_{i_n i_{\sigma(n)}^\prime} \delta_{j_1 j_{\sigma(1)}^\prime} \ldots \delta_{j_n j_{\sigma(n)}^\prime} + \cdots .
\end{equation}
Using this approximation,  we can easily obtain the $n$-th Rényi entropy of the radiation region.
We can describe the index contraction  in $\tr(\rho^n)$  by using a diagrammatic method as shown in Fig. \ref{fig:v}. After averaging under the Haar measure, we obtain an integral over the Haar measure multiplied by a coefficient. By calculating these, we can derive the expression for the entropy. The details of the computation is as follows. First, we can simplify the diagram, as shown in Fig.  \ref{fig:simp}. Since $|\mathrm{X}\rangle_{rR} |\mathrm{X^\prime}\rangle_{r^\prime R^\prime}$ represents the maximally entangled state in the regions $rr'$ and $RR'$, it can be written as 
\begin{equation}    
|\mathrm{X}\rangle_{rR} |\mathrm{X^\prime}\rangle_{r^\prime R^\prime}= \sum_{x=1}^{|X|} \frac{1}{\sqrt{|X|}}|x\rangle_{rr'}|x\rangle_{RR'}.\end{equation} 
For the index contraction on the region $RR'$, the entanglement is transferred to the contraction of the indices on $rr'$ through the maximally entangled state, and the factors $\frac{1}{\sqrt{|X|}}$ on both sides of the contraction matrix are retained. Thus, after simplifying the schematic diagram to the index contraction on the region $rr'$, we need to multiply it by an additional factor of $\frac{1}{|X|}$.

\begin{figure}[t]
	\centering
	\includegraphics[width=1.0\linewidth]{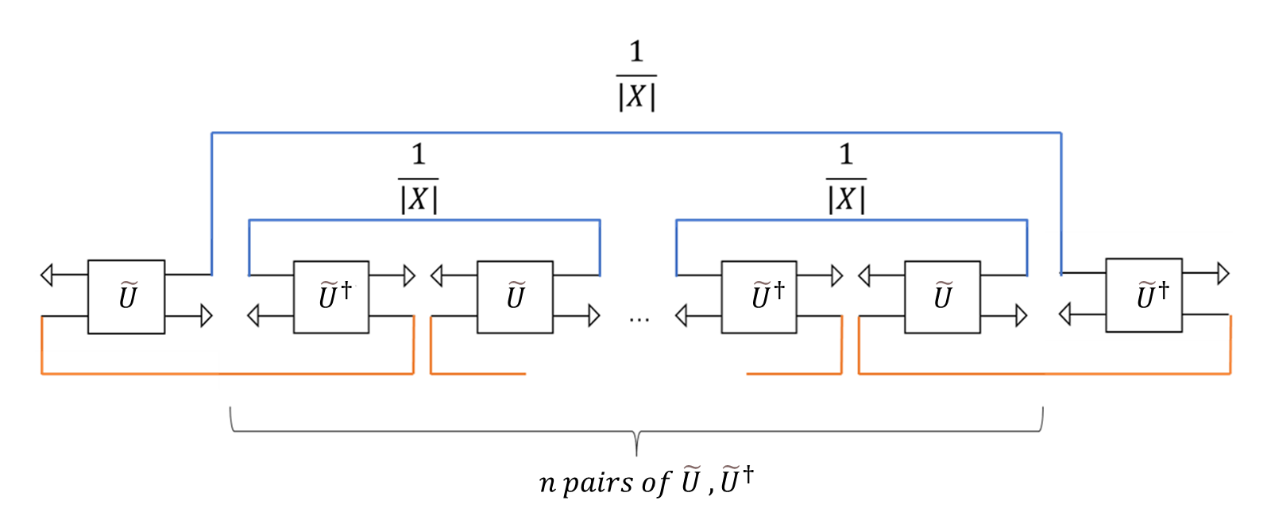}
	\caption{The diagram illustrating the calculation of the $n$-th Rényi entropy. The red lines represent the contraction of indices over the regions $B$ and $B^\prime$. After contracting the red lines, we obtain the density matrix of the radiation region, $\rho_{RR^\prime}(\tilde{U})$. The blue lines represent the contraction of indices over the regions $R$ and $R^\prime$. We replicate the density matrix $n$ times, and the contraction of the blue lines represents the operation of multiplying the $n$ copies of the density matrix and taking the trace, where the indices $x$ of $\tilde{U}^\dag$ are cyclically permuted.}
	\label{fig:entropyrr}
\end{figure}

After simplifying the diagram, we can more conveniently describe the calculation process of the $n$-th Rényi entropy of the radiation, whose density matrix is:
\begin{equation}
	\rho_{RR^\prime}(\tilde{U}) = \tr_{BB^\prime}\left(  \tilde{V} \otimes I_{RR^\prime} |\psi\rangle \langle \psi| \tilde{V}^\dag \otimes I_{RR^\prime} \right).
\end{equation}
As shown in Fig. \ref{fig:entropyrr}, we replicate the system $n$ times, resulting in $n$ pairs of $\tilde{U}$ and $\tilde{U}^\dag$. From equation (\ref{Vtilmap}), for each $\tilde{V}$ or $\tilde{V}^\dag$, the normalization factor is $|X|$, so the total normalization factor is $|X|^{2n}$. In Fig. \ref{fig:entropyrr} we have $n$ factors of $\frac{1}{|X|}$ from $n$ index contractions, so the whole coefficient  should be $|X|^n$. The Haar integral can be expressed as
\begin{equation}
	\int d\tilde{U} \prod_{i=1}^n \tilde{U}_{y_i x_i} \tilde{U}^\dag_{x_{\tau(i)} y_i} = \frac{1}{N^n} \sum_{\sigma \in S_n} \prod_{i=1}^n \delta_{x_i x_{\sigma \tau(i)}} \delta_{y_i y_{\sigma(i)}} = \frac{1}{N^n} \sum_{\sigma \in S_n} |X|^{|\sigma \circ \tau|} Y^{|\sigma|}\label{Haarintegral}
\end{equation}
where $x$ and $y$ correspond to the indices of regions $rr'$ and $BB'$, respectively, $i$ labels the $i$-th copy; $\tau$ and $\sigma$ belong to the elements of permutation group $S_n$: $\tau$ is the cyclic permutation and $\sigma$ is any group element; $|\sigma|$ represents the number of cycles in the permutation $\sigma$.

We can generally write the expression for the $n$-th  Rényi entropy as
\begin{equation} S^{(n)} = -\frac{1}{n-1}\log\left( \sum_{\sigma \in S_n} f_n(\sigma) \right).
\end{equation} 
When we find the maximum value $f_n^{\text{max}}$ in $f_n(\sigma)$, we have 
\begin{equation} -\frac{1}{n-1}\log(f_n^{\text{max}}) - \frac{1}{n-1}\log(\Gamma(n+1)) \leq S^{(n)} \leq -\frac{1}{n-1}\log(f_n^{\text{max}}) .
\end{equation} 
Fitting as $n$ tends to 1, we see that \begin{equation} -\lim_{n \rightarrow 1} \frac{1}{n-1} \log(f_n^{\text{max}}) - (1 - \gamma) \leq S^{(1)}\leq -\lim_{n \rightarrow 1} \frac{1}{n-1} \log(f_n^{\text{max}}) 
\end{equation}
where $\gamma = 0.577...$, i.e. 
\begin{equation} S^{(1)} = -\lim_{n \rightarrow 1} \frac{1}{n-1} \log(f_n^{\text{max}}) + O(1). 
\end{equation}
In the summation \eqref{Haarintegral}, because $|X|$ and $|Y|$ are much larger than 1, we only need to find the dominant contribution.

According to \cite{Akers:2021pvd}, we have $|\sigma| + |\sigma \circ \tau| \leq n + 1$. Thus, when $|X| \ge |Y|$, the term with the largest $|\sigma \circ \tau|$ dominates. In this case, $\sigma = \tau^{-1}$, so $|\sigma \circ \tau| = |e| = n$ and $|\sigma| = |\tau^{-1}| = 1$. Multiplying by the coefficient $|X|^n$, the expression for the Rényi entropy in this case is
\begin{equation}\label{XY1}
	\int d\tilde{U} \tr(\rho_{RR^\prime}^n(\tilde{U})) = \frac{1}{|Y|^{n-1}}.
\end{equation}
When $|X| \le |Y|$, the term with the largest $|\sigma|$  dominates. In this case, $\sigma = e$, so $|\sigma \circ \tau| = |\tau| = 1$ and $|\sigma| = |e| = n$, where $e$ is the identity element. Multiplying by the factor $|X|^n$, the expression for the Rényi entropy in this case is
\begin{equation}\label{XY2}
	\int d\tilde{U} \tr(\rho_{RR^\prime}^n(\tilde{U}))= \frac{1}{|X|^{n-1}}.
\end{equation}
Thus, the  fine-grained entropy of the radiation is
\begin{equation}\label{SRR}
	S(RR') = 2 \min(\log |R|, \log |B|),
\end{equation}
 which agrees with the result given by the QES formula.

\section{Reflected entropy}\label{sec:rfl}

In this section, we would like to compute the reflected entropies between different regions in the two-sided black hole model.
First, let us review the definition of the reflected entropy \cite{Dutta:2019gen}. For a subsystem $AC$, its Hilbert space can be factorized into a direct product of two sub-Hilbert spaces $\mathcal{H}_{AC} = \mathcal{H}_A \otimes \mathcal{H}_C$. The density matrix on $AC$ is $\rho_{AC}$. It can be canonically purified  on a double Hilbert space $\mathcal{H}_A \otimes \mathcal{H}_C \otimes \mathcal{H}_{A^\ast} \otimes \mathcal{H}_{C^\ast}$ to $|\sqrt{\rho_{AC}}\rangle$ with
\begin{equation}\label{puri}
	|\sqrt{\rho_{AC}}\rangle = \rho_{AC} \otimes I_{A^\ast C^\ast} |\phi_{AC}^+\rangle
\end{equation}
where $I_{A^\ast C^\ast}$ is the identity matrix and $|\phi_{AC}^+\rangle$ is a maximally entangled state without normalization factor, $\sum_i |i\rangle_{AC} |i\rangle_{A^\ast C^\ast}$. The reflected entropy between $A$ and $C$ is defined as
\begin{equation}
	S_R(A:C
	) \equiv S({AA^\ast})_{\sqrt{\rho_{AC}}} = S(\rho_{AA^\ast})
\end{equation}
where $\rho_{AA^\ast} = \text{tr}_{CC^\ast}(|\sqrt{\rho_{AC}}\rangle \langle \sqrt{\rho_{AC}}|)$ is the reduced density matrix of the $AA^\ast$ subsystem.

Based on the strong subadditivity of the entanglement entropy we can determine the upper and lower bounds for the reflected entropy:
\begin{equation}
	I(A:C)\le S_R\left(A:C\right)\le2\min(S\left(A\right),S\left(C\right)).
\end{equation}
In the following, in addition to calculating the reflected entropy, we also calculate the mutual information.  Actually, we may define the so-called Markov gap\cite{Hayden:2021gno}, 
\begin{equation}
   S_R\left(A:C\right)- I(A:C),
\end{equation}
which can be related to the fidelity of a particular Markov recovery process on the canonical purification of $\rho_{AC}$. 

As the computation of the entanglement entropy, one may calculate the $n$-th Rényi reflected entropy first and then do analytical continuation $n\to 1$ to obtain the reflected entropy. However, unlike the usual entanglement entropy, the calculation of the reflected entropy involves the operator $\sqrt{\rho_{AC}}$, which  is not easy to define and compute. To solve this problem, one introduce a new replication on $\sqrt{\rho_{AC}}$ \cite{Dutta:2019gen} to define a normalized state,
\begin{equation}
	|\phi^{(m)}\ran= \tr \left( \rho_{AC}^m\right)^{-1/2} |\rho_{AC}^{m/2}\ran.
\end{equation}
When $m$ approaches 1, we return to our initially defined purification. When $m$ is even, the terms containing $\sqrt{\rho_{AC}}$ do not appear, making the calculation easy. At the same time, considering $n$ copies for the $n$-th Rényi entropy, one obtain a $(m, n)$-th Rényi entropy, and the reflected von Neumann entropy is given by
\begin{equation}
	S_R(A:C
	) = \lim_{m,n \to 1} \frac{1}{1-n} \log \text{tr} \left( \text{tr}_{CC^\ast} \left( |\phi^{(m)} \rangle \langle \phi^{(m)}| \right)^n \right).
\end{equation}

In the following calculation, it is necessary to take the average over the Haar measure. To prevent the denominator from containing the unitary matrix random variable, we should first take average on the normalized denominator
\begin{equation}
	|\overline{\phi^{(m)}}\ran= \tr \left(\overline{ \rho_{AC}^m}\right)^{-1/2} |\rho_{AC}^{m/2}\ran.
\end{equation}
The modified definition of reflected entropy is
\begin{equation}
S_R(A:C
) = \lim_{m,n \to 1} \frac{1}{1-n} \log \text{tr}\left(\overline{ \text{tr}_{CC^\ast} \left( |\overline{\phi^{(m)}}\ran \lan\overline{\phi^{(m)}}| \right)^n }\right). 
\end{equation}
 In this definition, when averaging over the Haar measure, we can conveniently apply the Haar measure integration formula (\ref{approxint}). 

\subsection{Reflective entropy between black holes}

In this subsection, our goal is to calculate the reflected entropy $S_R(B: B^\prime)$ and the mutual information $I(B: B^\prime)$ between the  black hole regions on two sides.

\begin{figure}
	\centering
	\includegraphics[width=01.0\linewidth]{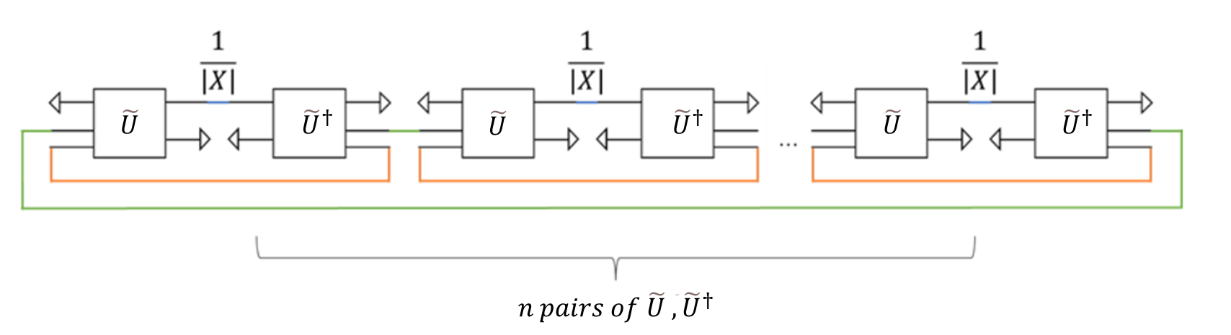}
	\caption{The diagram showing how to compute the $n$-th power of  the reduced density matrix of region $B$.  In this diagram, the blue line represents the simplified index contraction multiplied by a factor of $\frac{1}{|X|}$, the red line represents the index contraction for the region $B^\prime$, and the green line represents the index contraction and taking trace of the $n$ replicated reduced density matrices of $B$. Using the indices of $\tilde{U}$ as the basis, among the indices of $\tilde{U}^\dag$, the red line corresponds to the $B^\prime$ and the blue line corresponds to the $X$ with the same indices as $\tilde{U}$, while the green line corresponds to the indices of $B$ with a cyclic permutation of the $n$ elements.}
	\label{fig:entropyb}
\end{figure}

First, we calculate the mutual information between the two black hole regions $B$ and $B^\prime$, given by
\begin{equation}
	I(B: B^\prime)= S(B) + S(B^\prime) - S(BB^\prime).
\end{equation}
Since the entire system in our model is in a pure state, the entanglement entropy of the two-sided black hole region $S(BB^\prime)$ should be equal to the one of the two-sided radiation region $S(RR^\prime)$. According to (\ref{SRR}), we obtain
\begin{equation}
	S(BB^\prime)= 2 \min(\log |R|, \log |B|).
\end{equation}
Due to symmetry, we  expect that $S(B)$ is equal to $S(B^\prime)$. Therefore, the only quantity we need to calculate is the entanglement entropy corresponding to the region $B$. The reduced density matrix of $B$ is
\begin{equation}
	\rho_B(\tilde{U}) = \tr_{RR^\prime B^\prime} \left( \tilde{V} \otimes I_{RR^\prime} |\psi\rangle \langle \psi| \tilde{V}^\dag \otimes I_{RR^\prime} \right)
\end{equation}
We can calculate the $n$-th Rényi entropy  of the region $B$. From Fig. \ref{fig:entropyb}, similar to the previous analysis, we can factor out a coefficient $\left|X\right|^n$, and for the Haar integral we have
\begin{equation}
	\int d\tilde{U} \prod_{i=1}^n \tilde{U}_{B_i B_i'; x_i} \tilde{U}^\dag_{x_i; B_{\tau(i)} B^\prime_i} = \frac{1}{N^n} \sum_{\sigma \in S_n} \prod_{i=1}^n \delta_{B_i B_{\sigma(\tau(i))}} \delta_{B_i' B'_{\sigma(i)}} \delta_{x_i x_{\sigma(i)}} = \frac{1}{N^n} \sum_{\sigma \in S_n} |B|^{|\sigma \circ \tau|+|\sigma|} |X|^{|\sigma|}.
\end{equation}
Since $\left|\sigma \circ \tau\right| + \left|\sigma\right| \le n + 1$, the equality holds when $\sigma$ is the identity element, and $\left|\sigma\right| = \left|e\right| = n$ takes the maximum value. Thus, the $n$-th Rényi entropy of $B$ satisfies
\begin{equation}
	\int dU \text{tr}(\rho^n_B(\tilde{U})) = \frac{1}{|B|^{n-1}}. \quad 
\end{equation}
The von Neumann entropy of $B$ is simply
\begin{equation}
	S(B) = \log |B|.
\end{equation}
Thus we can obtain an expression for the mutual information based on the above results,
\begin{equation}
	I(B:B')=2\max(\log|B|-\log|R|,0).
\end{equation} 
Therefore, we see that the quantum correlation between the  black holes gets weakened by the Hawking radiations.  After the Page time, there is no longer correlation between the two black holes, and all the degrees of freedom in the  black holes are entangled with $R$ and $R'$.

Now we turn to  the calculation of the reflected entropy $S_R(B:B')$. For the purification defined in (\ref{puri}), its effect on the external legs is merely to rotate the directions of the external legs by 180 degrees (from the perspective of the diagram), meaning that  the vector corresponding to the external leg is turned to its dual vector. Because the maximally entangled state $|\phi^+\rangle$ does not have a normalization factor, we do not need to add any additional factors before the expression. Now we consider the $m$-replicated state $|\phi^{(m)}\rangle$. We can first calculate its denominator $\tr(\overline{\rho_{BB^\prime}^m})^{1/2}$, which after averaging becomes a coefficient outside the Haar integral. To read the $m$-th Rényi entropy of the two-sided black hole region $BB'$, we need to compute $\tr(\overline{\rho_{BB^\prime}^m})$ and have
\begin{equation}
	\tr(\overline{\rho_{BB^\prime}^m}) = \max\left(\frac{1}{|X|^{m-1}}, \frac{1}{|Y|^{m-1}}\right).
\end{equation}
To compute the reflected entropy, we need to consider the $(m,n)$-th Rényi entropy by using double-replica trick, which requires us to replicate the density matrix $n$ times, so the denominator, when multiplied $n$ times, contributes a factor of $\left(\max\left(\frac{1}{|X|^{m-1}}, \frac{1}{|Y|^{m-1}}\right)\right)^{-n}$. 

\begin{figure}
	\centering
	\includegraphics[width=01.0\linewidth]{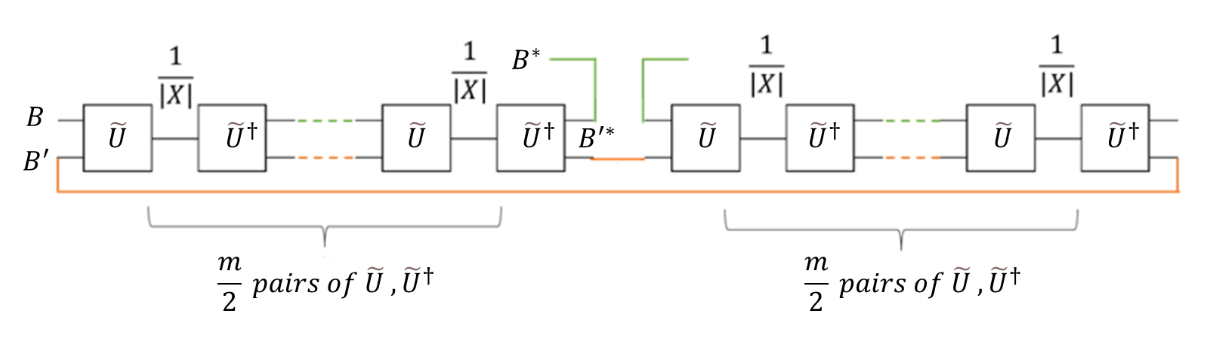}
	\caption{The diagram showing the computation of the numerator of the reduced density matrix $\tr_{B'B^{\prime\ast}}(|\rho_{BB'}^{m/2}\rangle \langle \rho_{BB'}^{m/2}|)$,  for even $m$ replicated states. The fixed external legs were omitted as they do not affect the calculation. We have labeled the regions corresponding to each external leg. From the green solid lines, it can be seen that the conjugate transpose of the degrees of freedom corresponding to $B$ becomes the degrees of freedom for $B^\ast$.}
	\label{fig:reduced-density}
\end{figure}

\begin{figure}
	\centering
	\includegraphics[width=1.0\linewidth]{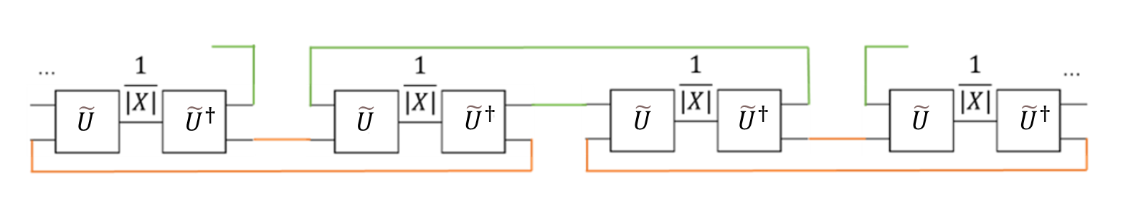}
	\caption{When $m=2$, the diagram for the numerator of the reduced density matrix. The red lines and green lines represent the index contractions in regions $B'$ and $B$ respectively, while the black lines represent the index contractions in $x$, accompanied by a factor of $\frac{1}{|X|}$. From the cyclic units, it is easy to see that if the indices of $\tilde{U}$ are fixed, the indices $x$ of $\tilde{U}^\dag$ correspond one-to-one with the indices $x$ of $\tilde{U}$. The indices $B$ and $B'$ of $\tilde{U}^\dag$ are composed of a series of length $m$ cycles, arranged in an interleaved manner. The end of one cycle of one index is located in the middle of the cycle of the other index. From this, the expressions for $g_1$ and $g_2$ can be read.}
	\label{fig:m2ref}
\end{figure}

After handling the coefficient contributed by the denominator, we next need to calculate the numerator of the reduced density matrix $\tr_{B'B^{\prime\ast}}(|\rho_{BB'}^{m/2}\rangle \langle \rho_{BB'}^{m/2}|)$. We refer to the copied region for the two-sided black hole region $BB'$ as $B^\ast B'^\ast$.  Actually, to compute the reflected R\'enyi entropy, $m$ should be even.  In this case, we can use the  diagram shown in Fig. \ref{fig:reduced-density}. We have a total of $m$ pairs of $\tilde{U}$ and $\tilde{U}^\dag$ such that  the reduced density matrix  corresponding to  $|\phi^{(m)}\rangle$  gives rise to a factor of $\frac{1}{|X|^m}$. Taking into account the fact that  computing  the $(m,n)$-th Rényi entropy  requires replicating $n$ times, we find a factor of $\frac{1}{|X|^{mn}}$. 
Meanwhile, the normalization factor for $nm$ pairs of $\tilde{U}$ and $\tilde{U}^\dag$ is $|X|^{2mn}$. Therefore, before calculating the Haar integral, we should multiply it by an overall coefficient 
\begin{equation}
    |X|^{mn}\left(\max\left(\frac{1}{|X|^{m-1}}, \frac{1}{|Y|^{m-1}}\right)\right)^{-n}
    \end{equation}
The relevant  Haar integral is
\begin{equation}\label{refBB}
\int d\tilde{U}\prod_{i=1}^{nm}{\tilde{U}_{B_iB_i';x_i}{\tilde{U}^\dag}_{x_i;B_{g_2\left(i\right)}B_{g_1\left(i\right)}'}}=\frac{1}{N^{mn}}\sum_{\sigma\in S_n}{\left|B\right|^{\left|\sigma\circ g_1\right|+\left|\sigma\circ g_2\right|}\left|X\right|^{\left|\sigma\right|}},
\end{equation}
where $g_1$ and $g_2$ are two group elements in the permutation group $S_n$, satisfying
\begin{equation}
	g_1 = (1, 2, \ldots, m)\ldots(n-m+1, \ldots, n),
\end{equation}
\begin{equation}
	g_2 =\left(\frac{m}{2}+1,\ldots\frac{3m}{2}\right)\ldots\left(\left(n-\frac{1}{2}\right)m+1,\ldots,nm,1,\ldots\frac{m}{2}\right).
\end{equation}
We  show the calculation of $(2, n)$-th reflected Rényi entropy  in Fig. \ref{fig:m2ref}.

When $|X| \ge |Y|$, namely the dimension of the radiation region is larger than that of the black hole region, corresponding to the late evaporation stage,  the dominant term in (\ref{refBB})  is the one that $\left|\sigma\right|$ takes the maximum value. In this case, $\sigma$ must be the identity element $e$ with $\left|\sigma\right| = nm$, and $\left|\sigma \circ g_1\right| + \left|\sigma \circ g_2\right| = 2n$. Thus, multiplying by the factor $|X|^{mn} |Y|^{mn-n}$, we obtain the reflected  entropy:
\begin{equation}
	S_R(B:B') = 0.
\end{equation}
This indicates that in the late evaporation stage, the reflected entropy $S_R(B:B')$ reaches its lower bound, simply vanishing.

When $|X| \le |Y|$, namely the dimension of the radiation region is smaller than that of the black hole region, corresponding to the early evaporation stage, the situation is more complex. We need to introduce the Cayley distance\cite{Akers:2021pvd}, defined by the minimal number of transpositions. For the permutation group $S_n$, let $\alpha, \beta$ be group elements, then we have the Cayley distance
\begin{equation}
	d(\alpha, \beta) = n - \left|\alpha \circ \beta^{-1}\right|.
\end{equation}
The Cayley distance satisfies the triangle inequality,
\begin{equation}
	d(\alpha, \beta) + d(\beta, \gamma) \geq d(\alpha, \gamma).
\end{equation}
Thus, we obtain
\begin{equation}
	\left|\sigma \circ g_1\right| + \left|\sigma \circ g_2\right| \leq 2nm - 2n + 2.
\end{equation}
If $|Y|$ is large enough, we need to first consider the exponent of $|Y|$. In order to get the dominant term in (\ref{refBB}), we need the inequality to be saturated. In terms of the Cayley distance this requires that the $\sigma$ should be a permutation of the  minimal number of transposition from $g_1$ to $g_2$:
\begin{equation}
    d(g_1, \sigma^{-1}) + d(\sigma^{-1}, g_2) = d(g_1, g_2).
\end{equation}
 We find that $g_1$ and $g_2$ differ  by integer multiples of $m/2$, from $1$ to $nm$. As $g_1^{-1}g_2$ is the transposition from $g_1$ to $g_2$, if we  omit the identical maps, $g_1^{-1}g_2$ can be written as 
\begin{equation}
g_1^{-1}g_2=\left(
\begin{array}{lr}
m, \frac{m}{2}, 2m, \frac{3m}{2}, \ldots, nm, (n-\frac{1}{2})m &\\

	nm, \frac{3m}{2}, m, \frac{5m}{2}, \ldots, (n-1)m, \frac{m}{2} &
\end{array}
\right).
\end{equation}
In this case, $\sigma^{-1}$ belongs to one of the minimal number of transpositions, namely  $\sigma^{-1}$ is in an intermediate process of the cycle of $(m, 2m, \ldots, nm)$ and $(m/2, 3m/2, (n-1/2)m)$. Then we need to choose a $\sigma$ with maximal $\left|\sigma\right|$ in these intermediate processes. By induction we find that when $\sigma = g_1^{-1}$ or $\sigma = g_2^{-1}$, $\left|\sigma\right|$ has a maximum value $n$. Thus, multiplying by the factor $|X|^{2mn-n}$, we obtain the reflected  entropy:
\begin{equation}\label{SRBB}
	S_R(B:B') = 2 \log |B| .
\end{equation}
But if $\log|X|$ and $\log|Y|$ are comparable, when the exponent  $\frac{1}{2}(\left|\sigma \circ g_1\right| + \left|\sigma \circ g_2\right|)$ of $|Y|$ decreases and the exponent $|\sigma|$ of $|X|$ increases, $\left|B\right|^{\left|\sigma\circ g_1\right|+\left|\sigma\circ g_2\right|}\left|X\right|^{\left|\sigma\right|}$ may increase. Considering $\frac{1}{2}(\left|\sigma \circ g_1\right| + \left|\sigma \circ g_2\right|)+|\sigma|\le nm+2n$, when 
$\frac{1}{2}(\left|\sigma \circ g_1\right| + \left|\sigma \circ g_2\right|$)
 decreases by $1$ from its maximum value, 
$|\sigma|$ can increase by at most 
$n$, which happens  when
\begin{equation}
    \sigma=(1,2,...\frac{m}{2})^{-1}(\frac{m}{2}+1,\frac{m}{2}+2...m)^{-1}...((n-\frac{1}{2})m+1,...,nm)^{-1}.
\end{equation}
So strictly speaking, the derivation to \eqref{SRBB} is valid only when $|X|^n\le |Y|$. However, if we take $n\rightarrow1$, this requirement goes back to $|X|\le |Y|$. In the following computations, we will use the same treatment. 

Considering the upper bound of the reflected entropy
\begin{equation}
	2 \min(S(B), S(B')) = 2 \log |B|, 
\end{equation}
 we see that when the dimension of the radiation region is smaller than that of the black hole region (early evaporation), the reflected entropy $S_R(B:B')$ equals its upper bound $2 \log |B|$. On the contrary, when the dimension of the radiation region is larger than that of the black hole region (late evaporation), the reflected entropy $S_R(B:B')$ equals its lower bound, 0.

In the late stage of evaporation, the fact that the entanglement entropy of the radiation region equals the black hole  entropy $2 \log |B|$  suggests that at this time, the black hole region and the radiation region are maximally entangled, and there is no entanglement between the two black hole regions. Hence, the reflected entropy is zero. This  picture has also been  verified  in the two-sided black hole model in which  the JT gravity is coupled with a conformal field theory\cite{Li:2020ceg}. In fact, if we take the high-temperature limit $\beta\rightarrow0$, the reflected entropy Eq. (5.19) in \cite{Li:2020ceg}
\begin{equation} 
    S_{R}(B_L : B_R) = 2S_0 + 2\Phi_r +  \frac{c}{3}(b- \ln \cosh t + \ln 2)
\end{equation}
can be approximated as
\begin{equation} 
    S_{R}(B_L : B_R) = \frac{2\pi c}{3\beta}(b- t).
\end{equation}
The single-side black hole entropy Eq.(D.1) in \cite{Li:2020ceg}
\begin{equation}
    S(B_L)=S_0 + \frac{\phi_r}{\tanh d}  + \frac{c}{6}  \ln \frac{4\sinh^2 \frac{b+d}{2}}{\sinh d} 
\end{equation}
can be approximated as
\begin{equation} 
    S(B_L)= \frac{\pi c}{3\beta}b. 
\end{equation}
So before the evaporation the reflected entropy between the two black holes in the  JT gravity+CFT
 model $S_{R}(B_L : B_R)$ also equals the upper bound $2S(B_L)$.

In the early stage of evaporation, although there is an entanglement entropy $2 \log |R|$ in the radiation region, the dimension of the black hole region is larger than that of the radiation region. Therefore, the two black holes are approximately in a pure state, and the reflected entropy is saturated at the upper bound $2 \log |B|$. It is worth noting that in our model, the contribution of the entanglement between the black hole region and the radiation region to the reflected entropy is of higher order. This also leads to a discontinuous  transition of the reflected entropy at the Page time,  while the  transition of the Page curve of the entanglement entropy only has a discontinuity in the first derivative at the Page time.

\subsection{Reflective entropy between  radiations}

\begin{figure}
	\centering
	\includegraphics[width=0.5\linewidth]{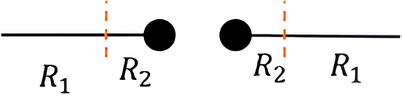}
	\caption{For the partition of the two-sided radiation region, the red dashed line divides the radiation region on one side into two sub-regions $R_1$ and $R_2$. Both of these two sub-regions are disconnected but are symmetrically distributed in the radiation region $R$ and its copy. In our model, due to the symmetry of the unitary matrix $\tilde{U}$, other less symmetric partitions  might yield the same result.
	}
	\label{fig:cut}
\end{figure}

 Here  we are not going to study the  reflected entropy between the radiations on two sides. Instead,  we partition the whole radiation region in a way similar to that employed in \cite{Li:2020ceg}.   Specifically, we cut the radiation region and its copy at the same positions to obtain two  regions $R_1$ and $R_2$, as shown in Fig. \ref{fig:cut}. Note that both sub-regions $R_1$ and $R_2$ are disconnected and distributed in an symmetric way.

Similar to the previous subsection, we first compute the mutual information $I(R_1:R_2)$, assuming without loss of generality that $|R_1|>|R_2|$. In the mutual information
\begin{equation}
I(R_1:R_2) = S(R_1) + S(R_2) - S(R_1R_2),
\end{equation}
 $S(R_1R_2)$ is the von Neumann entropy of the whole radiation region $R$, given by
\begin{equation}
S(R_1R_2) = 2\min(\log |R|, \log|B|) .
\end{equation}
Next, we need to compute $S(R_1)$ and $S(R_2)$. Similar to the previous calculation of R\'enyi entropy, the key lies in identifying the index correspondence in the Haar integral and determining the dominant elements in the permutation group $S_n$. Analogous to the previous calculations, the coefficient is $\left|X\right|^n$, and the expression for the Haar integral is
\begin{equation}
\int d\tilde{U}\prod_{i=1}^{nm} \tilde{U}_{y_i;R_{1i}R_{2i}} \tilde{U}_{R_{1\tau(i)} R_{2i}:y_i} = \frac{1}{N^n} \sum_{\sigma \in S_n} |Y|^{|\sigma|} |R_2|^ {|\sigma|} |R_1|^{|\sigma\circ\tau|}.
\end{equation}
When $\left|R_1\right|\ge\left|R_2\right|\left|Y\right|$, $\sigma$ should be  taken as the inverse of an $n$-cycle permutation $\tau^{-1}$ such that $\left|\sigma\circ\tau\right|=n$. As $\left|\sigma\right|=1$, and multiplying by $\left|X\right|^n$, we find the $n$-th Rényi entropy and further obtain the von Neumann entropy of the region $R_1$,
\begin{equation}
	S(R_1) = 2\log |B| + \log |R_2|.
\end{equation}
When $\left|R_1\right|\le\left|R_2\right|\left|Y\right|$, $\sigma$ should be the identity element $e$, $\left|\sigma\circ\tau\right|=1$, $\left|\sigma\right|=n$, and multiplying by $\left|X\right|^n$ gives the $n$-th Rényi entropy. Then, we obtain the von Neumann entropy of the region $R_1$,
\begin{equation}
S(R_1) = \log|R_1| .
\end{equation}
From the above discussion, we have
\begin{equation}
S(R_1) = \min(\log |R_1|, 2\log |B| + \log |R_2|).
\end{equation}
Symmetrically, we obtain the expression for the von Neumann entropy of region $R_2$:
\begin{equation}
S(R_2) = \min(\log |R_2|, 2\log |B| + \log |R_1|) = \log |R_2|.
\end{equation}
Finally, we have the mutual information $I(R_1:R_2)$ between the regions $R_1$ and $R_2$ with $|R_1|>|R_2|$
\begin{equation}
I(R_1:R_2)=S(R_1)+\log |R_2|-2\min(\log |R|, \log|B|)
\end{equation}
It can be classified into  three stages, with two  transition points. The first transition occurs at $\left|R\right|=|B|$, which corresponds to the Page time. The second transition occurs at $\left|R_1\right|=\left|B\right|^2|R_2|$, which was called the Page time for the region $R_1$ \cite{Li:2020ceg}. Before the first transition, the mutual information $I(R_1:R_2)=0$; between the first and second transitions, $I(R_1:R_2)=2\log{\left|R\right|-2\log{\left|B\right|}}$; and after the second transition, $I(R_1:R_2)=2\log{\left|R_2\right|}$. In summary, we have ($|R_1|>|R_2|$)
\begin{equation}
I(R_1:R_2)=\left\{\begin{array}{ll}
0,&\mbox{when $|R_1||R_2|\leq |B|$}\\
2\log{\left|R\right|-2\log{\left|B\right|}},&\mbox{when $|B||R_2|^{-1}\leq |R_1|\leq \left|B\right|^2|R_2|$}\\
2\log{\left|R_2\right|},&\mbox{when $|R_1|\geq \left|B\right|^2|R_2|$}.
\end{array}
\right.
\end{equation}
This three-stage behavior in the mutual information also appears in Fig. 5.7 of \cite{Li:2020ceg}.

Next, we consider the reflected entropy between  the region $R_1$ and the region $R_2$. Similar to the previous calculations, we first compute the denominator. Since the total system is in a pure state, the value of the denominator is the same as in the previous section. We then  obtain the Haar integral multiplied by the coefficient (considering $(m,n)$-th Rényi entropy)
\begin{equation}
    |X|^{mn}\left(\max\left(\frac{1}{|X|^{m-1}}, \frac{1}{|Y|^{m-1}}\right)\right)^{-n}.
\end{equation}
Next, we identify the elements of the permutation group $S_n$ corresponding to the ways in contracting  the $nm$ replicated external legs. Due to the asymmetry between the region $R_1$ and the region $R_2$, the Haar integral corresponding to the $(m,n)$-th Rényi reflected entropy between the regions $R_1$ and $R_2$ differs slightly from the Haar integral corresponding to the $(m,n)$-th\ Rényi reflected entropy of the black holes $BB'$. Finally, we read  
\begin{equation}
	\int d\tilde{U}\prod_{i=1}^{nm} \tilde{U}_{y_i;R_{1i}R_{2i}} \tilde{U}^\dagger_{R_{1g_2(i)}R_{2g_1(i)};y_i} = \frac{1}{N^{mn}} \sum_{\sigma \in S_n} |R_1|^{|\sigma \circ g_2|} |R_2|^{|\sigma \circ g_1|} |Y|^{|\sigma| }.
\end{equation}
When $\left|Y\right| \ge |X|$, $\sigma$ should be taken as the identity element. The dominant term is $\left|R_1\right|^n \left|R_2\right|^n \left|Y\right|^{nm} = \left|X\right|^n \left|Y\right|^{nm}$, resulting in the $(m,n)$-th Rényi entropy being zero. Thus, we obtain the reflected  entropy between the region $R_1$ and the region $R_2$:
\begin{equation}
		S_R(R_1:R_2) = 0.
\end{equation}
When $\left|Y\right| \le |X|$, given $\left|R_1\right| > \left|R_2\right|$, $\sigma$ should be  taken as the inverse of $g_2$ to ensure the exponent of $\left|R_1\right|$ reaches its maximum. The dominant term is $\left|R_1\right|^{nm} \left|R_2\right|^{nm-2n+2} \left|Y\right|^{nm} = \left|X\right|^{nm} \left|Y\right|^n \left|R_2\right|^{-2n+2}$. We obtain the reflected  entropy between the region $R_1$ and the region $R_2$:
\begin{equation}
	S_R(R_1:R_2) = 2 \log|R_2|.
\end{equation}
Meanwhile, we can compute the upper bound of the reflected entropy $S_R(R_1:R_2)$:
\begin{equation}
	2 \min(S(R_1), S(R_2)) = 2S(R_2) = 2 \log|R_2|.
\end{equation}
Based on the above results , we find that before the Page time, the reflected entropy $S_{R}(R_1:R_2)$ is at its lower bound, and after the Page time, the reflected entropy $S_R(R_1:R_2)$ is at its upper bound. In short, we have 
\begin{equation}
    S_R(R_1:R_2)=\left\{\begin{array}{ll}
    0,& \mbox{when $|R_1||R_2|\leq |B|$}\\
    2 \mbox{min}(\log |R_1|, \log|R_2|), &\mbox{when $|R_1||R_2|> |B|$.}
    \end{array}\right.
\end{equation}
The result here is the same as the one in \cite{Akers:2022max}, where the reflected entropy in West Coast model was studied. 

There is only one  transition for the reflected entropy between the region $R_1$ and the region $R_2$, and no  transition at the Page time for $R_1$. Additionally, under the partition of the regions $R_1$ and $R_2$, the  transition of the reflected entropy $	S_R(R_1:R_2)$ is discontinuous, whereas the two  transitions of the mutual information $I(R_1:R_2)$ are continuous. 

\subsection{Radiation and black hole}

For an evaporating single-sided black hole, its fundamental description involves  the entanglement between a boundary quantum system $B$ and a single-sided radiation system $R$. In the case of a two-sided black hole coupled with a two-sided radiation system, the fundamental description is composed of the region of the original single-sided black hole $BR$ and its mirror copy $B'R'$ so that  the whole system gets canonically purified. This purification process conforms to the definition of reflected entropy, thus we should have
\begin{equation}\label{equ}
	S^{(s)}_R(R:B) = S(RR').
\end{equation}
The reflected entropy $S^{(s)}_R(R:B)$ between the radiation subregion $R$ and the black hole subregion $B$ in the single-sided evaporating black hole system, should be equal to the entanglement entropy of the two-sided radiation region $RR'$ in the two-sided evaporating black hole system.

However, it is worth noting that in our model, the purification is defined in the effective description, while the entropy calculation is performed by using the non-isometric mapping  in the  fundamental description. There is no direct quantitative relationship between the holographic mapping corresponding to the single-sided black hole and the one  corresponding to the two-sided black hole, i.e., $\tilde{V} \neq V \otimes V$, but we can verify that for  the holographic mapping described in this  work, the relationship in equation (\ref{equ}) still holds. 

\begin{figure}
	\centering
	\includegraphics[width=0.5\linewidth]{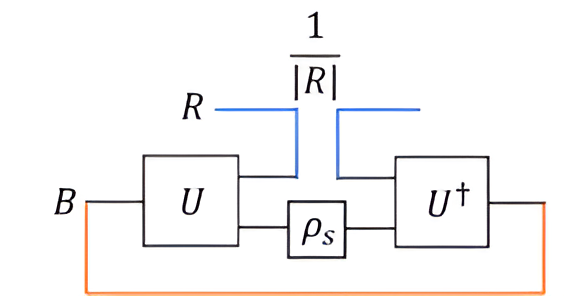}
	\caption{The diagram of the reduced density matrix corresponding to the radiation region in the single-sided evaporating black hole model. The three indices $B$, $R$, and $s$ that need to be summed are all marked in the figure. The red lines represent the tracing out of the single-sided black hole region when calculating the reduced density matrix. The blue curves represent the contraction of the unitary matrices $U$ and $U^\dag$ with the maximally entangled states on $r$ and $R$.} 
	\label{fig:rhos}
\end{figure}

 In order to calculate the reflected entropy $S^{(s)}_R\left(R:B\right)$ between the radiation subregion $R$ and the black hole subregion $B$ in the single-sided evaporating black hole model,  we first consider the properties of the reduced density matrix corresponding to a single radiation region $R$, as shown in Fig. \ref{fig:rhos}, where $U$ and $U^\dag$ are defined by (\ref{V}).  By simplifying the maximally entangled state to the external leg representing the dual vector, the normalization factor of the two maximally entangled states provides a factor of $\frac{1}{\left|R\right|}$.
After knowing the form of the reduced density matrix,  we are able to  calculate the $(m,n)$-th Rényi entropy. For the denominator, we have\cite{Akers:2022qdl}
\begin{equation}
	\tr(\overline{\rho_{BR}^m})= \max \left( \frac{1}{|R|^{m-1}},\frac{\tr(\rho_s^m)}{|B|^{m-1}} \right). 
\end{equation}
Combined with the previous discussion, a pair of $U$ and $U^\dagger$ corresponds to a coefficient of $\frac{1}{|R|}$, and from  (\ref{V}), the normalization coefficient for a pair of $U$ and $U^\dagger$ is $\left| R \right|^2$. As a result,  we get the total coefficient
\begin{equation}
 \left| R \right|^{mn} \max \left( \frac{1}{|R|^{m-1}},\frac{\tr(\rho_s^m)}{|B|^{m-1}} \right)^{-n} .  
\end{equation}  
We can also calculate the form of the Haar measure integration as
\begin{equation}
	\int dU\prod_{i=1}^{nm} U_{B_i;R_{i}a_{i}}{\rho_s}_{a_ib_i} U^\dagger_{R_{g_2(i)}b_{i};B_{g_1(i)}} = \frac{1}{N^{mn}} \sum_{\sigma \in S_n} (|R|^{|\sigma \circ g_2|} |B|^{|\sigma \circ g_1|} \prod_{r=1}^{|\sigma| }\tr(\rho_s^{c_r(\sigma)})),
\end{equation}
where $a,b$ are the indices of $\rho_s$, and $c_r$ is the length of the $r$-th cycle in the group element $\sigma$, satisfying $\sum_{r=1}^{|\sigma|}c_r(\sigma)=n$. According to the properties of the density matrix,  the dominant term is obtained when $\sigma = g_1^{-1}$ or $\sigma = g_2^{-1}$.
When $|R| \le |B|$, the dominant contribution  is $\left| R \right|^{nm-2n+2} \left| B \right|^{nm} \tr(\rho_s^m)^n$. In this case,  the reflected von Neumann entropy is
\begin{equation}
	S^{(s)}_R (R:B) = 2 \log |R|.
\end{equation}
When $|R| \ge |B|$, the dominant contribution  is $\left| R \right|^{nm} \left| B \right|^{nm-2n+2} \tr(\rho_s^m)^n$ and  the reflected von Neumann entropy is
\begin{equation}
	S^{(s)}_R (R:B) = 2 \log |B|.
\end{equation}
Thus we have
\begin{equation}
	S^{(s)}_R (R:B) = 2 \min (\log |R|, \log |B|) = S(RR'). 
\end{equation}
That is, in the single-sided evaporating black hole model, the reflected entropy $S^{(s)}_R (R:B)$ between the radiation subregion $R$ and the black hole subregion $B$ is equal to the entanglement entropy $S(RR')$ of the two-sided radiation in the two-sided evaporating black hole model.

Going a step further, if we purify multiple small single-sided black holes and perform the replacement $V \otimes V \rightarrow \tilde{V}$ multiple times, we can study the case of multi-sided black holes. By applying the previous computational techniques, we can study the multi-entropy and reflected entropy on evaporating multi-sided black holes. (Some previous studies can be found in\cite{Bao:2019zqc,Chu:2019etd,Yuan:2024yfg}.)

\section{Conclusion and discussion}\label{sec:con}
In this paper, we introduced  a two-sided evaporating black hole model by using a post-selection map defined by a Haar random unitary matrix. We investigated the entanglement entropy and reflected entropy in this model through a non-isometric map. The  entropy of radiation in the  model matches the result from the quantum extremal surface formula and agrees with the Page curve. 

In this model, we are allowed to discuss the reflected entropy and the related mutual information between different regions, including the black holes on different sides, the radiations  distributed in a symmetric but disconnected way, and the black hole and radiation on single side. The main results are as follows. 
\begin{itemize}
    \item For the reflected entropy between the black holes, it is at its upper bound before the Page time and at its lower bound (zero) after the Page time. The transition at the critical point is discontinuous, similar to the results in \cite{Li:2020ceg}.
    \item For the reflected entropy between the radiations, it is at its lower bound before the Page time and at its upper bound after the Page time, coincides with the Page curve of reflected entropy proposed in \cite{Akers:2022max}. The transition at the critical point is discontinuous. In contrast,  the corresponding mutual information present a three-stage behavior, undergoing the continuous transitions at the Page time and at the Page time for the region $R_1$. 
    \item Finally, we verified that in the single-sided evaporating black hole model, the reflected entropy $S_R\left(R:B\right)$ between the radiation subregion $R$ and the black hole subregion $B$ is equal to the entanglement entropy $S(RR^\prime)$ of the two-sided evaporating black hole model's radiation.
\end{itemize}

Our investigations rely on  the non-isometric maps from the effective description to  fundamental description. Even though the results in this work are consistent with the ones in other models\cite{Li:2020ceg,Akers:2022max}, our method is different from the ones in \cite{Li:2020ceg,Chandrasekaran:2020qtn,Akers:2022max}, which based on the semi-classical effective description. It would be interesting to study the reflected entanglement spectrum more carefully and investigate the possible correction to the above picture. Furthermore, our model can also be extended to the case of multi-sided black holes. In future work, we may be able to use our method to explore the multipartite generalization of reflected entropy from a different perspective. 


\section*{Acknowledgments}
    We are grateful to Zezhou Hu for useful discussions.  This research is supported in part by {NSFC Grant  No. 11735001, 12275004.}  \par
    
    \vspace{2cm}
    
\providecommand{\href}[2]{#2}\begingroup\raggedright\endgroup

\end{document}